# Two-dimensional flexible high diffusive spin circuits

*I. G. Serrano[1]\*, J. Panda[1]\*, Fernand Denoel[1]\*, Örjan Vallin[2], Dibya Phuyal[1], Olof Karis[1], and M. Venkata Kamalakar[1]*

[1]Department of Physics and Astronomy, Uppsala University, Box 516, SE-751 20 Uppsala, Sweden
[2]Department of Engineering Sciences, Uppsala University, Box 534, SE-751 21 Uppsala, Sweden

\*authors contributed equally
Email to: venkata.mutta@physics.uu.se

Owing to their unprecedented electronic properties, graphene and two-dimensional (2D) crystals have brought fresh opportunities for advances in planar spintronic devices. Graphene is an ideal medium for spin transport while also being an exceptionally resilient material for flexible electronics. However, these extraordinary traits have never been combined to create flexible graphene spin circuits. Realizing such circuits could lead to bendable strain-based spin sensors, a unique platform to explore pure spin current based operations and low power flexible nanoelectronics. Here, we demonstrate graphene spin circuits on flexible substrates for the first time. These circuits, realized using chemical vapour deposited (CVD) graphene, exhibit large spin diffusion coefficients ~0.19-0.24 $m^2s^{-1}$ at room temperature. Compared to conventional devices of graphene on $Si/SiO_2$ substrates, such values are 10-20 times larger and result in a maximum spin diffusion length ~10 µm in graphene achieved on such industry standard substrates, showing one order enhanced room temperature non-local spin signals. These devices exhibit state of the art spin diffusion, arising out of a distinct substrate topography that facilitates efficient spin transport, leading to a scalable, high-performance platform towards flexible 2D spintronics. Our innovation unlocks a new domain for the exploration of strain-dependent spin phenomena, and paves the way for flexible graphene spin memory-logic units and surface mountable sensors.

Electron spin physics holds vital ingredients for creating new kinds of energy-efficient electronic memory and logic components[1,2], that can accelerate computing, and promises additional means for further power saving through novel integration architectures[3]. The two-dimensional (2D) crystal, graphene, is a special material where spin-polarized electrons can travel tens of microns without losing their spin orientation at room temperature[4,5]. This extraordinary ability, stemming from its low spin-orbit coupling[6] and a negligible hyperfine interaction, brought a colossal drive for true planar multi-terminal spin circuits that are promising for enabling novel technologies beyond the conventional two-terminal magnetoresistance applications. Over the past decade, graphene spintronics has witnessed significant developments[7–9]. These include the initial demonstration of micrometer scale lateral spin transport[10], promising efficient spin transport[11], improved ferromagnetic interfaces with graphene[12] for efficient spin injection, higher quality graphene channels via complex encapsulated structures[8,13–15], dry transfer methods[16,17], and suspended schemes[18,19]. Developments also include spin relaxation studies[9,20–22], more recent anisotropic effects[23,24], observation of long-distance spin transport[4,5], novel enhancements[25,26], inversion effects using 2D insulators[27–30] as barriers for spin injection and innovative demonstrations[31,32]. While the majority of these experiments have been performed using small-scale flakes that are mechanically exfoliated from bulk graphite, developments using chemical vapour deposited (CVD) graphene have also revealed the feasibility of ballistic transport[33] and wafer-scale spintronic capability[5,33–36]. Despite all these developments and a rapidly evolving landscape, spin transport in graphene on flexible substrates has never been realized. The fact that 2D crystals can sustain extraordinary levels of strain makes them the best materials for flexible nanoelectronics[37]. At the same time, harnessing the unmatched spin transport capability of graphene can lead to 2D flexible spin circuits. Realizing such devices will have an impact for next-generation ultra-low power flexible spin-based nanoelectronics, prospective heatless processors, and novel sensors. In addition, such systems could lead to the discovery of new spin-strain correlations in strained 2D materials and their heterostructures with graphene. However, the challenges related to the fabrication methods, processing with flexible substrates, and concerns about efficiency have made the feasibility of flexible graphene spin circuits (FGSC) an open question. In particular, conventional research is oriented towards high-mobility (mobility ≥10 000 $cm^2s^{-1}V^{-1}$) heterostructures of graphene on atomically flat boron nitride substrates[13,14], heterostructures fabricated via dry transfer methods[16,17], and suspended graphene[18,19] devices. In contrast to such schemes that are aimed to circumvent substrate roughness-induced carrier scattering, industrial flexible substrates such as polyethylene-based derivatives possess nanometre levels of roughness[38], much higher than the roughness of even standard $Si/SiO_2$ wafer substrates. Apart from this obvious topographic disadvantage, the difficulties in nanofabrication of spin devices, realizing spin transport, and performing low-temperature experiments with thermal management in different measurement geometries are non-trivial, unlike the experiments involving devices on conventional substrates. Therefore, despite a decade-long progress in 2D flexible nanoelectronics[37] and high-quality graphene flexible field-effect devices[39–41], the feasibility and demonstration of 2D flexible spin devices remains a great challenge.

In an effort to make a generic conceptual advancement towards 2D flexible spin devices, to open gateways for exploring new spin phenomena and associated applications, here, we overcome the experimental challenges to demonstrate flexible graphene spin circuits (FGSC) for the first time. We have fabricated graphene lateral spin transport devices on polyethylene naphthalate (PEN) using large-scale CVD graphene. In such devices, through comprehensive electrical and spin transport experiments, we realized efficient spin transport in long graphene channels (up to 15 μm) at room temperature. Our experiments reveal a high diffusive spin transport with 20-fold enhancement of spin diffusion coefficient when compared to standard values for graphene devices on $Si/SiO_2$. This leads to one order enhanced non-local spin signal amplitudes, a long spin diffusion length ~10 μm, which is the highest value obtained in any form of graphene on such standard large-scale substrates. We discuss to what extent our results compare with the present state of the art results obtained in high-quality graphene devices. In addition, we explain how the nature and unique topography of the polyethylene naphthalate substrate leads to high spin diffusion in graphene, and what makes FGSC a versatile large-scale high-performance flexible platform for new advances in 2D spintronics.

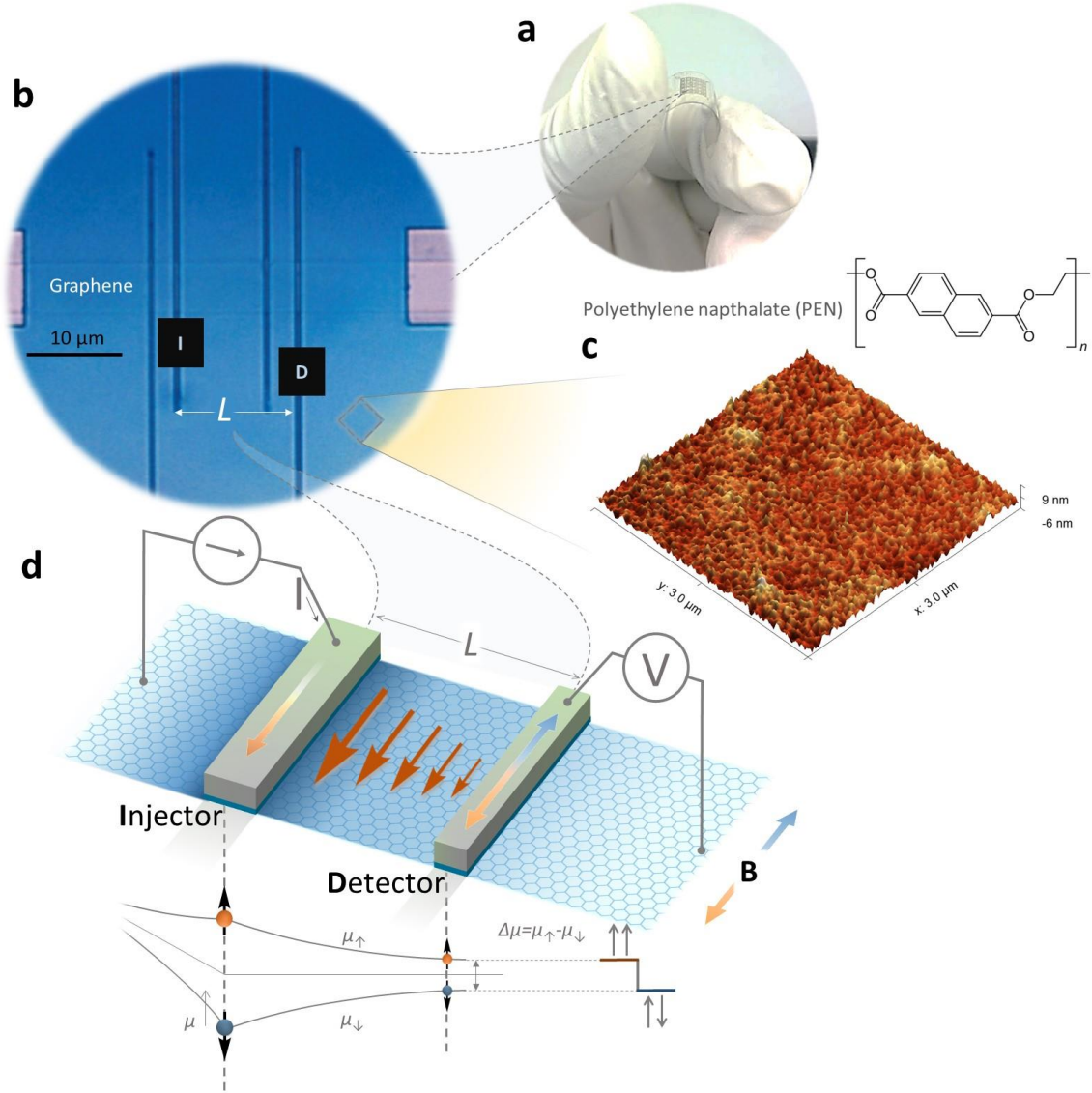

**Figure 1| Graphene lateral spin valves on flexible substrates. a,** Arrays of graphene spin devices fabricated on flexible PEN substrate. **b,** An optical micrograph of a FGSC with multiple contacts for electrical and spin transport characterization. **c,** Atomic force microscope image displaying the standard topography of the flexible substrate. **d,** Non-local measurement scheme for the detection of pure spin signals (pure spin currents are depicted by the arrows on the graphene channel). Below the device structure, the varying profile of the spin-up ($\mu_\uparrow$) and spin down ($\mu_\downarrow$) chemical potentials across the channel is shown. The current loop (I) performs electrical spin injection into graphene. The voltage (V) circuit measures the spin accumulation $\Delta\mu=\mu_\uparrow-\mu_\downarrow$ by switching the magnetic orientation of the injector and detector from parallel (↑↑ or ↓↓) to antiparallel (↓↑ or ↑↓) by applying an in-plane magnetic field (**B$_\parallel$**).

To fabricate FGSC, we started with a layer of large-scale CVD graphene grown over a high purity copper substrate. The large-scale nature of CVD graphene helps in creating arrays of devices with promise for high throughput, uniformity, and scalability. By an improved wet chemical processing technique[5], we transferred graphene on to flexible substrates. Through optimized processes of optical lithography, oxygen plasma etching, electron beam lithography assisted patterning, electron beam metal deposition, and lift-off methods, we have fabricated graphene spin-valve devices on the flexible PEN substrates. The PEN substrate is an industry standard semi-crystalline flexible substrate[38]. In relation to other widely used polyethylene derivatives such as polyethylene terephthalate (PET), it possesses improved processing temperature limit (150 °C) and a relatively lower topographic root mean square roughness ~1.3 nm (shown in Fig. 1c). It exhibits negligible shrinkage if the fabrication of the device is performed at a temperature less than 150°C and under optimized fabrication conditions[38]. However, in contrast to fabricating devices on Si/SiO$_2$, nanofabrication on flexible substrates requires significant changes such as optimizations of the exposure dose and other experimental conditions. The details of the fabrication process are explained in the Methods section and Supplementary information. In Fig. 1a, we display a photograph of an array of FGSC devices. An individual graphene spin circuit, as shown in Fig. 1b, consists of a patterned graphene channel, contacted by a series of titanium barrier-based cobalt ferromagnetic (2nm Ti|Co 65nm|10nm Au) electrodes. These contacts have different widths rendering each of them a distinct magnetic coercivity, which allows for switching of the magnetization of the individual electrodes at distinct values of the applied magnetic field. To perform electrical and spin measurements, the fabricated devices were connected with the help of a specially developed micro-bonding tip using high conductivity wires, ensuring necessary thermal anchoring of the samples for temperature dependent measurements. We performed detailed charge and spin transport experiments in the temperature range 77-300 K in a N$_2$/He flow cryostat to characterize the contacts, graphene channels, and demonstrate spin transport in the non-local scheme of pure spin signal measurements (Fig. 1d).

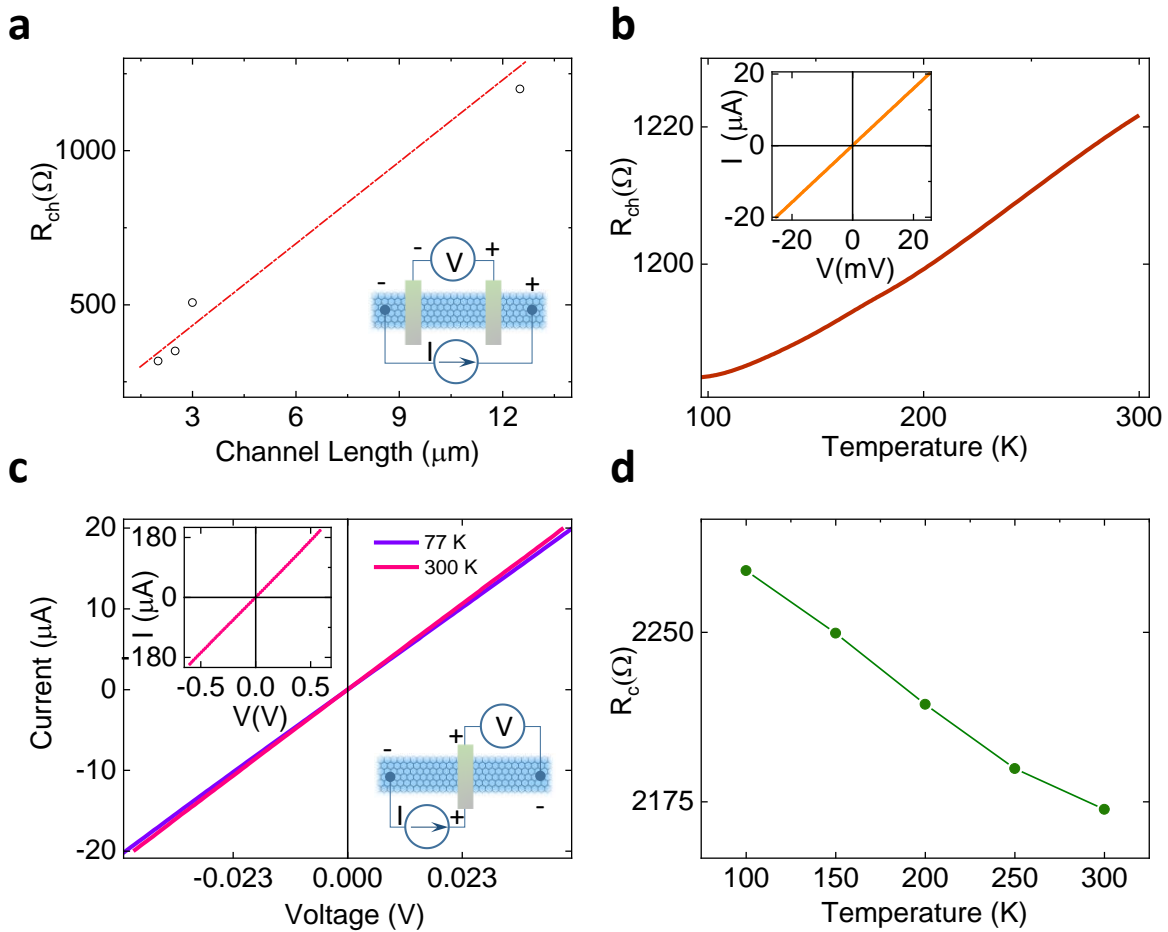

**Figure 2| Electrical characterization of graphene channels and contacts. a,** Four-probe electrical resistance plotted for different channel lengths displaying a linear scaling (denoted by the dotted line)

of the channel resistance with the length of the channel. The inset shows the four-probe measurement configuration. **b,** Four probe resistance of a graphene channel as a function of temperature. Inset: four-probe current-voltage (IV) characteristic of the channel. **c,** IV characteristics of a typical contact in the device at different temperatures. The insets show a high bias IV characteristic plot and a standard three terminal measurement configuration. **d,** Temperature dependence of the three-terminal contact resistance.

Prior to performing spin transport experiments, it is important to evaluate the electrical nature of the individual contacts and graphene channel, both of which are essential to determine the feasibility of spin transport. In Fig. 2a, we show the resistance values measured in a standard four-probe measurement technique for different channel lengths of the device. We observed a linear dependence of the resistance with channel length, leading to channel square resistance ~400 Ω/sq., which indicates a reasonably good and uniform electronic quality of graphene stripes up to a length of 30 μm. The temperature-dependent channel resistance, displayed in Fig. 2b, exhibits a typical electron-phonon scattering-dominated behaviour, as observed for graphene on Si/SiO$_2$ and polymethyl methacrylate (PMMA) substrates[42,43]. The spin injection/detection contacts in our devices showed a typical interface contact resistance $R_c$~2000 Ω (resistance area ~ 2000 Ω μm$^2$ with contact area ~ 1 μm$^2$), with linear IV characteristics, as presented in Fig. 2c. The inset of Fig. 2c shows a schematic of the standard three-terminal measurement configuration used for these measurements. We observed a weak temperature dependence of the contact resistance (as seen in Fig. 2d), with less than 4% change in the entire measurement range (77-300 K). In contrast to high resistive tunnelling contacts of aluminium oxide[10,44], magnesium oxide[12,17] having resistance area ~10-100 kΩ μm$^2$, here the contacts are Ohmic over a relatively high bias voltage (up to ± 0.5 V as shown in the inset of Fig .2c), and exhibit significantly low resistance. The values are broadly closer to expected values for direct titanium contacts on graphene[45]. Achieving spin transport in graphene with low resistance contacts could help in reducing power dissipation during electrical spin injection. Despite the relatively low resistance, the values of the contact resistance for both the injector and detector are higher than the graphene channel resistance. Such contacts satisfy the necessary criteria for spin devices overcoming the conductivity mismatch problem[46,47], and fit into the optimum range for spin injection into the graphene channels (shown in Fig. S2, supplementary information section 2). Having characterized the electrical nature of the contacts and channel, we now focus on the spin transport experiments. In a lateral graphene spin device, the quantification of pure spin currents in graphene is performed through a non-local (NL) measurement scheme represented in Fig. 1d. In the NL scheme, the spin currents are electrically injected into the graphene channel through the current circuit (*I*), and the resulting diffusion spin current at a distance *L* is detected by the voltage circuit (*V*). Such an isolation of the current and voltage circuits allows for faithful measurement of pure spin accumulation, eliminating spurious magneto-resistive contributions associated with direct charge currents. By sweeping an in-plane magnetic field (**B$_∥$**), the relative magnetic orientations of the injector and detector electrodes are switched from parallel to anti-parallel configuration which results in a spin-valve switching in the measured non-local spin voltage (**ΔV$_{NL}$**= V$_{NL}$(↓↑)- V$_{NL}$(↑↑)). As depicted in the spin chemical potential profile (Fig. 1d), this allows for the measurement of the difference in spin chemical potentials for up (μ$_↑$) and down spins (μ$_↓$), also known as the net spin accumulation at a distance *L*. In our experiments, by performing forward (**-B** to **+B**) and reverse (**+B** to **-B**) magnetic field sweeps, we could observe spin transport in long graphene channels. For a channel length *L* ≈12.5 μm , a clear spin valve signal obtained at 200 K is shown in Fig. 3a. The NL resistance ΔR$_{NL}$ = ΔV$_{NL}$/I = 800 mΩ at 200 K is significantly high for such a long distance. The evolution of the NL voltage in the temperature range 77-300 K, shown in Figs. 3b and 3c, reveals fairly large values ranging from 1 Ω at 77 K to 0.35 Ω at room temperature. Despite low resistive Ohmic contacts, the room temperature (300 K) value of ΔR$_{NL}$ = 0.35 Ω is nearly one order larger than what has been observed in large-scale graphene on silicon/SiO$_2$ substrates[5] for similar CVD graphene channel lengths. In fact, this magnitude compares well to spin valve amplitudes obtained in similar long spin transport channels of high-quality mechanically exfoliated graphene[4,23]. In the spin valve signal, a small slope is detected following the sharp switching of the shape anisotropy-guided magnetization of the electrodes. This could arise from a gradual orientation of some magnetic domains having slightly different orientation than the shape anisotropy or the applied magnetic field as observed in graphene spin valves earlier[32,48]. As seen in Fig. 3d, the bias dependence of the non-local spin voltage reveals a slight suppression of the spin signal for negative injection currents. Such non-linearity has been observed previously in graphene spin valves, and could possibly originate from a bias-dependent polarization of the spin injector and detectors[29]. These comprehensive measurements in the spin-valve geometry allow us to confirm robust spin transport in FGSC for the first time.

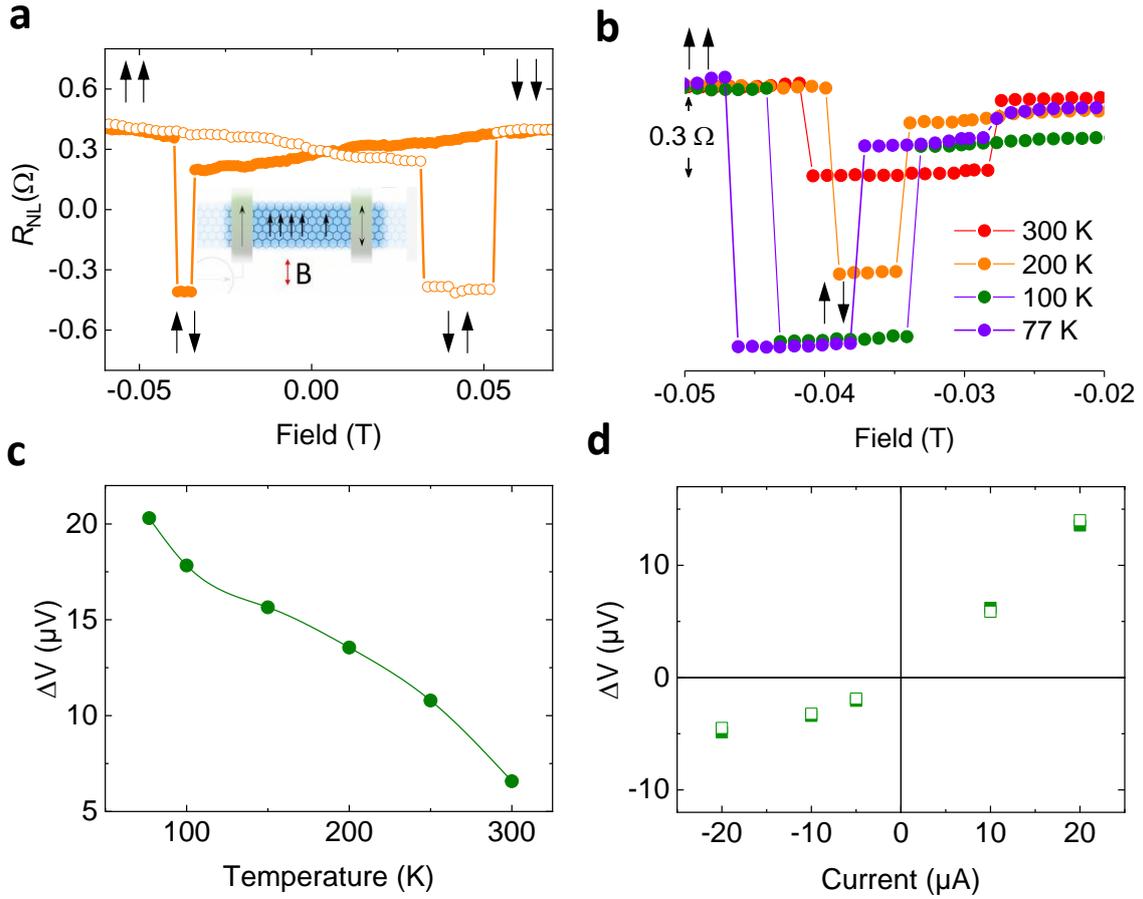

**Figure 3| Spin transport and precession through graphene in FGSC. a,** Spin transport signal measured in the spin valve geometry showing switching between parallel (↑↑ *or* ↓↓) and anti-parallel (↓↑ *or* ↑↓) configurations of the injector and detector electrodes; inset scheme represents the spin transport geometry. **b,** Spin valve switching signals ($R_{NL}$) obtained in the temperature range 77-300 K. **c,** Temperature dependence of the spin valve amplitudes. **d,** Bias dependence of the spin signal at 200 K (open and filled squares denote two switching amplitudes for two sweep directions).

To further quantify the spin transport in FGSC, we investigated the modulation of the spin signal as a function of an out-of-plane magnetic field ($B_\perp$), while maintaining a parallel configuration of the magnetization of the injector and detector electrodes. This scheme, also known as the Hanle measurement geometry, involves making the spin-polarized electrons in graphene undergo a Larmor precession with a frequency $\omega = \frac{g\mu_B}{\hbar} B_\perp$ (gyromagnetic ratio g = 2) about the out-of-plane magnetic field $B_\perp$. Such precession along with diffusion and spin relaxation contributions leads to a systematic modulation of the detected spin signal, which can be described by the solution to the steady-state Bloch diffusion equation

$$D\nabla^2 \mu_s - \frac{\mu_s}{\tau} + \omega \times \mu_s = 0 \qquad (1)$$

where $\mu_s = \mu_\uparrow - \mu_\downarrow$ is the net spin polarization vector with direction pointing the polarization axis, and $D$ and $\tau$ represent the spin diffusion constant and spin relaxation time, respectively, leading to a spin diffusion length $\lambda = \sqrt{D\tau}$. We fit the Hanle spin precessional signal to the analytical solution[49] $R_{NL}$ given by Eq. (2).

$$R_{NL} \propto Re\left\{ \frac{\sqrt{1+i\omega\tau}}{R_N} \frac{e^{-L/\sqrt{D\tau/(1+i\omega\tau)}}}{\left(1+\frac{2R_i}{R_N}\sqrt{1+i\omega\tau}\right)\left(1+\frac{2R_d}{R_N}\sqrt{1+i\omega\tau}\right) - e^{-2L/\sqrt{D\tau/(1+i\omega\tau)}}} \right\} \qquad (2)$$

where $L$ is the length of the channel, $R_N = \lambda R_{ch}/L$, $R_i$ and $R_d$ represent the injector and detector resistances respectively (in our devices the injector and detector have similar resistance values). Fig. 4a shows the spin precession signal near room temperature (RT) and at 77 K, both of which were fit to the Hanle Eq. 2 to extract the spin parameters. For an effective channel length $L \approx 12.5$ μm, we obtained a spin lifetime $\tau \approx 0.27$ ns with $D \approx 0.24$ m$^2$s$^{-1}$, yielding a spin diffusion length $\lambda \approx 8.04$ μm, which increases to 10 μm at 77 K. The obtained values of the diffusion constant $D$ in our experiments are up to 20 times larger than what has generally been observed in graphene on Si/SiO$_2$[5,9,10]. These values are also comparable to state of the art reports of high mobility graphene devices[4,13,14,16–19]. In addition to the results presented here for the first device (device-1 with sheet resistance ~400 Ω/square and low contact resistances), In Fig. 4b, we present the Hanle spin signal for a 15 μm-long channel device (device-2) obtained at room temperature. Fitted to the Hanle expression (Eq.2), device-2 yields a $\tau \approx 0.39$ ns and a $D \approx 0.19$ m$^2$s$^{-1}$, in a channel with a sheet resistance ~ 0.9 kΩ/sq and contact resistances ~7 kΩ. While the higher spin lifetime is in agreement with a higher contact resistance, the relatively lower $D$ (in comparison to that of device-1) obtained here could be ascribed to a relatively higher sheet resistance of the graphene channel. We note that while these devices were prepared following the same standardized fabrication route, the contacts showed different resistances which could be attributed to a possible unintentional partial oxidation of the contacts or batch-to-batch variation. Regardless of such variations, the magnitude of $D$ in our devices is one order higher than standard graphene spin devices on Si/SiO$_2$.

To scrutinize the spin diffusion in FGSC further, we performed Hall effect measurements and estimated a carrier hall mobility ~10 000-15 000 cm$^2$s$^{-1}$V$^{-1}$ (see supplementary information section 3). From measurements on hall devices prepared using the same CVD graphene sheet (with similar range of sheet resistance), we obtained the carrier mobility and calculated the charge diffusion coefficient ($D_c$). The calculated $D_c$~0.1 m$^2$s$^{-1}$ in our devices revealed a reasonable agreement with extracted spin diffusion coefficients, which is considered as a confirmation of the reliability of the Hanle fitting procedure[15,19]. Furthermore, in Fig. 4c, we also analyze the spin signal $\Delta R_{NL}$ vs. $L$ from measurement performed in channels lengths ~ 2.5-15 μm at room temperature. The $\Delta R_{NL}$ vs. $L$ plot shows a characteristic reduction in spin signal with increased channel lengths. A fit to the exponential decay of spin signal[5] with the length of channels of spin transport yields $\lambda \approx 8.2$ μm, similar to the obtained values for devices-1 and 2. The closer magnitude of high spin diffusion D in our devices, an order of magntiude enhanced spin signals, and the reasonable agreement with $D_c$ and the characteristic $\Delta R_{NL}$ vs. L yielding a similar diffusion length $\lambda$ indicate that the high $D$ in FGSC is significantly robust. These facts reveal that spin transport in FGSC is remarkably efficient, leading to a $\lambda \approx 10$ μm at low temperatures. To the best of our knowledge, the λ~8-8.6 μm is the highest room temperature value for any form of graphene (exfoliated, CVD, epitaxial) on standard large-scale substrates near ambient temperatures. Unlike such large values obtained in small-scale high mobility graphene complex heterosturctures[4,13,14,16,18,19], FGSC is a scalable system that has the dual novelty of being a flexible as well as possessing state of the art performance in graphene.

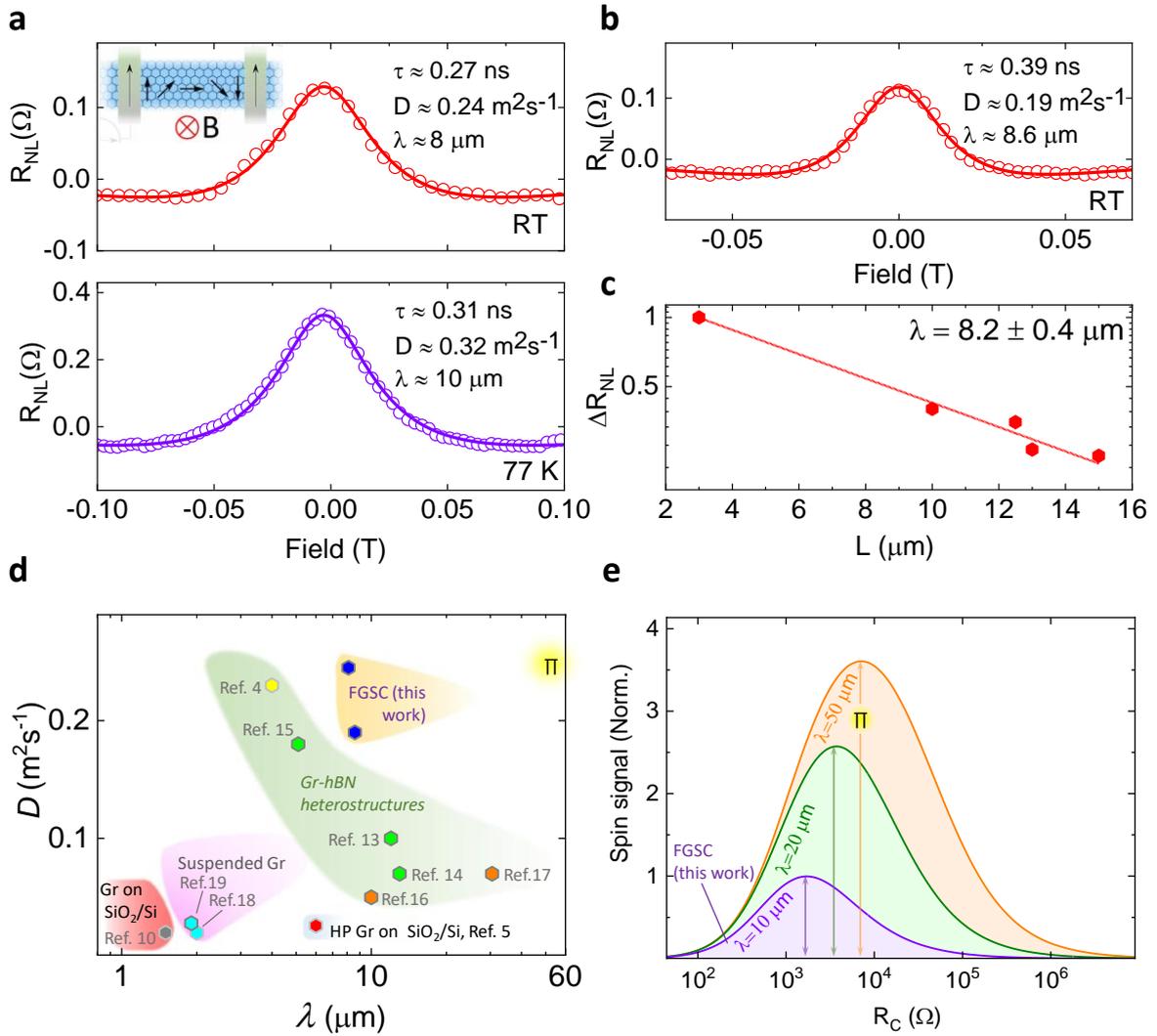

**Figure 4| Spin parameters and comparison with high-performance graphene spin-valve systems: a,** Hanle spin precession signals in a device with L ≈ 12.5 μm; inset scheme representing spin precession in graphene in an out-of-plane magnetic field. The solid lines are fits to the Hanle spin modulation (Eq. 2). **b,** Hanle data for a device with L ≈ 15 μm. **c,** Spin signal amplitude $\Delta R_{NL}$ vs. L for different channels at room temperature (RT). **d,** Spin diffusion constant ($D$) vs. spin diffusion length ($\lambda$) for different pristine (no intentional doping) graphene (Gr) uniform channels that have exhibited high performance (HP) near room temperature. The diffusion constant (D) has been calculated for Refs. 16, 17, and 18. **e,** Calculated spin signal vs. contact resistance obtained using Fert-Jaffrè[46] magneto-resistance calculations (see supplementary information section 2) for $\lambda$ = 10 μm, 20 μm, and 50 μm. The arrows indicate the corresponding contact resistance required for highest spin signals with such values of $\lambda$. The symbol ∏ indicates highest efficiency that can be achieved with highest D observed in our experiments, together with improvements in spin lifetime (described in the text).

In general, the intrinsic roughness of polymer substrates is not favourable for spin transport, and conventionally atomically flat substrates show improvements in spin parameters in graphene. In fact, the observed carrier mobility in our devices is up to one order of magnitude higher than the typical mobility in graphene on Si/SiO$_2$ substrate[5,50] and similar order of magnitude seen in some of the high mobility schemes [4,13,14,16,18,19] based-graphene spin devices. Although trapped charges or hydroxyl radicals can be low[51] in the case of soft polymeric substrates which could assist in improved mobility[39,40], a high substrate roughness may not be conducive for spin transport in graphene and the role of such roughness still poses a question. As a matter of fact, the PEN substrate with higher root mean square roughness ~1.3 nm (4-5 times higher than Si/SiO$_2$ roughness ~0.3 nm), provides a better spin transport than graphene over Si/SiO$_2$ systems and displays a state of the art spin diffusion D value. This is particularly counterintuitive as smoother substrates result in reduced modulation in substrate topography, thus, leading to reduced scattering potentials experienced by the charge carriers. To understand this apparent contradiction, we examined the topography of the flexible substrate using atomic force microscopy (AFM) (see supplementary information section 4 and Fig. S4). Interestingly, in PEN, despite a higher roughness, the roughness width (the peak-valley-peak distance) of ~50-100 nm is up to 5 times higher than that of SiO$_2$. A simple estimation reveals that this implies a potential reduction of up to 90% of the scattering peaks for a given area, in contrast to Si/SiO$_2$. In addition, the intra-peak roughness of PEN (≤1 Å) is also smaller than the intrinsic roughness of conventional Si/SiO$_2$ wafers. This unique feature is a new insight and gives a practical understanding of the reduction in channel-induced spin relaxation, and the observed high spin diffusive transport in FGSC.

Spin diffusion length $\lambda$ gives a direct measure of the distances over which spin communication can be established in a material. To highlight the significance of high $D$ observed here, in Fig. 4d, we plot $D$ vs. $\lambda$ values collected from literature reports of high-diffusive graphene channels. As a standard, we chose the first report of room temperature spin transport in graphene on Si/SiO$_2$ substrates[10] as a representative of spin transport parameters widely observed in graphene on Si/SiO$_2$. We also place the highest performance room temperature report[5] in CVD graphene on Si/SiO$_2$ as a benchmark. Next, we gather room temperature data that include high mobility exfoliated graphene on hexagonal boron nitride crystal[4], suspended graphene[19], graphene encapsulated in hexagonal boron nitride layers[13–15], and graphene-hexagonal boron nitride heterostructures prepared using dry transfer techniques[16,17]. It is worthwhile to pointout that a high charge carrier mobility does not always imply high spin diffusion, as it can still be challenging to obtain high spin diffusion[18,19]. In Fig.4d, it can be noticed that despite a relatively moderate spin lifetime due to low resistive contact, a high $\lambda$ can still be achieved due to state of the art $D$ values in FGSC.  One might argue from the figure that state of the art graphene-hexagonal boron nitride heterostructures still have the highest values of $\lambda$ due to higher spin lifetimes. However, the message here is that FGSC is an unconventional system that shows state of the art spin diffusion constant $D$ (0.19-0.24 m$^2$s$^{-1}$), in technologically attractive CVD graphene. It sets a new benchmark $\lambda$ in graphene over standard large-scale substrates at room temperature, while opening up a new branch of flexible graphene spintronics, and paves the way for practical graphene spintronics applications. It should also be noted that the high-performance achieved here is without any top coverage or encapsulation or employing dry transfer methods, which have shown to improve spin transport properties in graphene. We would like to mention here that higher spin diffusion values have been reported in graphene|WS2 composite channels[52], and at cryogenic temperatures in bilayer graphene[14]. However, for a fair comparison, here we have included only room temperature high-performance reports showing high spin diffusion in uniform graphene channels.

Our work unambiguously demonstrates the realization of flexible 2D spin circuits for the first time, where we also uncover a state of the art high spin diffusion in graphene. Beyond this first demonstration, as mentioned earlier, the spin diffusion length $\lambda \sim 10$ μm ($\lambda \approx 8.6$ μm at 300 K) is the highest value obtained in any form of graphene over large substrates. Here, we comment on some of the future prospects and potential that our work presents. As seen here, with enhanced R$_C$/R$_{ch}$, it is possible to obtain higher spin lifetime $\tau$ (as in the case of device-2 despite a higher sheet resistance). This is consistent to the findings that the spin absorption model (Eq.2) gives a more reliable value of $\tau$ accounting for the actual R$_C$ and R$_{ch}$ (unlike the assumption R$_C$ >> R$_{ch}$) [53], but the spin relaxation at the interface is mainly dominated by other contact-induced spin relaxation such as interface inhomogeneity fringe fields[53] rather than spin absorption. Therefore, this leaves a sound scope to mitigate the contact-induced spin relaxation[36,54] to raise spin lifetime to ~1-2 ns by interface engineering, for instance, with 2D insulators[26]. A high D obtained here multiplied with such improved spin lifetime could lead to spin diffusion lengths $\lambda$~25 μm in graphene. In addition to such enhancements, there is still scope for further enhancements in $\tau$ ~5-10 ns with top coverage and gate-assisted electrostatic doping[17]. Therefore, there is a real potential for FGSC to show $\lambda \geq 50$ μm. To check the device viability for such enhancements, in Fig. 4e, we discuss the calculated plots for spin signal vs. $R_C$ for several values of $\lambda$ using graphene sheet resistance of device 1. The optimum value $R_C$ ~10kΩ for highest efficiency, can be greater~20kΩ if we utilize graphene sheet properties of device-2. Such values of $R_C$ could be engineered using 2D insulators, as shown before[25,26]. It is noteworthy that with significantly low-contact resistance devices, we obtained $\lambda$ ~ 10 μm in graphene. Such low resistive contacts are important to reduce power dissipation at the contacts, enable high signal to noise ratio in spin sensors or even to enhance switching efficiency of spin-torque applications by effective spin absorption. Note that in our studies, all our devices were subjected to bending with a radius of curvature ~ 4 mm (as shown in Fig. 1a) prior to measurements. In-situ application of strain to study its influence on spin transport would be challenging but could enable discovering unseen strain-spin correlations and the impact of pseudo-magnetic fields[55], strain-controlled conductance[56], and unlocking new phenomena. In addition, FGSC presents a new means to investigate spin relaxation experiments in graphene on a topographically different landscape. Apart from such exciting opportunities, FGSC show a real promise for developing surface mountable graphene spin sensors, while the transparency of the substrates opens new avenues for novel investigations based on optical stimulus and applications. A large-scale high-performance graphene spintronic system is a basic need for exploring the possibility of low-power graphene spin logic and memory devices using pure spin current based operations and FGSC shows a unique promise in this direction.

In conclusion, we have been successful in realizing the first flexible graphene spin circuits and demonstrating state of the art spin diffusion in graphene. Through detailed experiments, we reveal diffusion constants ~10-20 times larger than that observed in conventional graphene devices on Si/SiO$_2$, and show one-order enhanced spin signals at room temperature. This enables the observation of spin diffusion lengths ~ 10 μm (8.6 μm at 300 K), highest achieved in graphene on industrial standard substrates. Our results present a flexible, scalable high-performance system that could emerge as a versatile platform for future 2D spintronic advancements. This innovation creates fresh prospects for a new realm of strain-spin phenomena, investigation of novel spin relaxation, and advancements towards pure spin current-based applications and bendable, transparent, low-power 2D flexible nanoelectronics.

**Methods:**
**Device Fabrication.** Flexible polyethylene naphthalate substrate obtained from DuPont was cleaned with acetone and isopropanol. A layer of commercially obtained CVD graphene was transferred onto the substrate using PMMA-assisted wet transfer process. Following the transfer, the substrate containing the CVD graphene layer was heated at 110 °C, before being transferred to a vacuum chamber for further drying. For processing into devices, the large-scale graphene was initially patterned into stripes of graphene micro-ribbons by optical lithography and controlled oxygen plasma etching. The oxygen plasma etching rate and power were calibrated to avoid unwanted damage to the substrate and resist hardening, and perform efficient etching of the graphene. Post etching, a well-optimized resist, and dose calibrated e-beam lithography process was used to pattern the electrodes. E-beam evaporation was used for deposition of a composite layer (2nm Ti|Co 65nm|10nm Au) consisting of 2 nm titanium seed layer, 65 nm Co, and a final 10 nm Au layer to protect the cobalt from oxidation, followed by lift-off in hot acetone. The typical electrodes have widths in the range 100-250 nm, and the effective contact area with graphene is ~0.5-1 µm$^2$. The resulting devices were imaged using an optical microscope, with the optical parameters adjusted to observe the contrast of graphene over PEN.

**Substrate and graphene characterization.** AFM was performed to assess the surface topography of the substrate. Raman Spectroscopy confirmed the single layer quality of CVD graphene (see Fig. S1.2). The channel resistance of the graphene sample was measured via four-probe measurement.

**Electrical and spin transport measurements.** To perform different electrical characterization and spin transport and precession measurements, the device electrodes were bonded using high-conductivity gold wire with a homemade micro-tip based bonder, compatible with soft substrates. Measurements were performed in a nitrogen/helium flow cryostat (77-300 K) with a room temperature electromagnet, using a Keithley source meter and nanovoltmeter. Interface resistance measurement of the contacts in three-terminal configuration and non-local measurements using four-terminal configuration were performed using currents in the range ±20 µA.

**Acknowledgments:** MVK acknowledges the VR starting Grant (No. 2016-03278) from the Swedish Research Council, funding from the Olle Engkvist foundation and the Wenner Gren Foundation. OK thanks funding from the Knut and Alice Wallenberg foundation. MVK thanks Ivan J. Vera-Marun for useful discussions.

**Author contributions:** IGS, MVK, JP, and FD fabricated and measured the devices reported here. MVK and OV optimized the device fabrication process with assistance from DP. MVK designed and set up the experiments with participation from OK. MVK conceived, supervised the project, analysed the results and wrote the manuscript with inputs from OK, IGS, JP, and FD. All authors have read and commented on the manuscript.

# Supplementary information

# Two-dimensional flexible high diffusive spin circuits

*I. G. Serrano[1]\*, J. Panda[1]\*, Fernand Denoel[1]\*, Örjan Vallin[2], Dibya Phuyal[1], Olof Karis[1], and M. Venkata Kamalakar[1]*

[1]Department of Physics and Astronomy, Uppsala University, Box 516, SE-751 20 Uppsala, Sweden
[2]Department of Engineering Sciences, Uppsala University, Box 534, SE-751 21 Uppsala, Sweden

\*authors contributed equally
Email to: venkata.mutta@physics.uu.se

1. **CVD graphene transfer and Raman spectroscopy**

Unlike in the case of normal silicon/SiO$_2$ substrates, transferring large-scale chemical vapour deposited (CVD) graphene on flexible substrates is non-trivial. The low optical contrast of graphene caused by the transparency of the flexible polyethylene napathlate (PEN) substrate also poses challenges during optimizations. In our process, the copper foil containing CVD graphene was attached to a heat release tape to ensure the mechanical stability necessary for spin coating of the polymethyl methacrylate (PMMA) resist layer. The copper from the Cu|graphene|PMMA stack was then etched out completely by a cupric chloride solution, rinsed in dilute hydrochloric acid, and washed with deionized water several times. Following this, we transferred the PMMA layer containing graphene to a PEN substrate. Post-transfer, the substrate with the film is dried, first by heating, and placed in a vacuum chamber prior to further processing. As it can be observed in Fig. S1.1, the PMMA edge in the middle of the frame is clearly visible. It is possible to distinguish the side of the substrate with PMMA from the clear side of the PEN without PMMA. Prior to nanofabrication, the PMMA layer was dissolved in hot acetone, followed by rinsing with isopropanol. The substrate containing the monolayer of CVD graphene was heated to 110 °C, following which nanofabrication was performed.

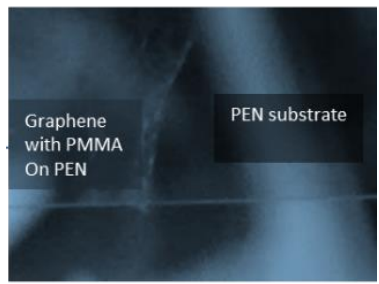

Fig. S1.1. A final dried layer of graphene|PMMA on PEN substrate.

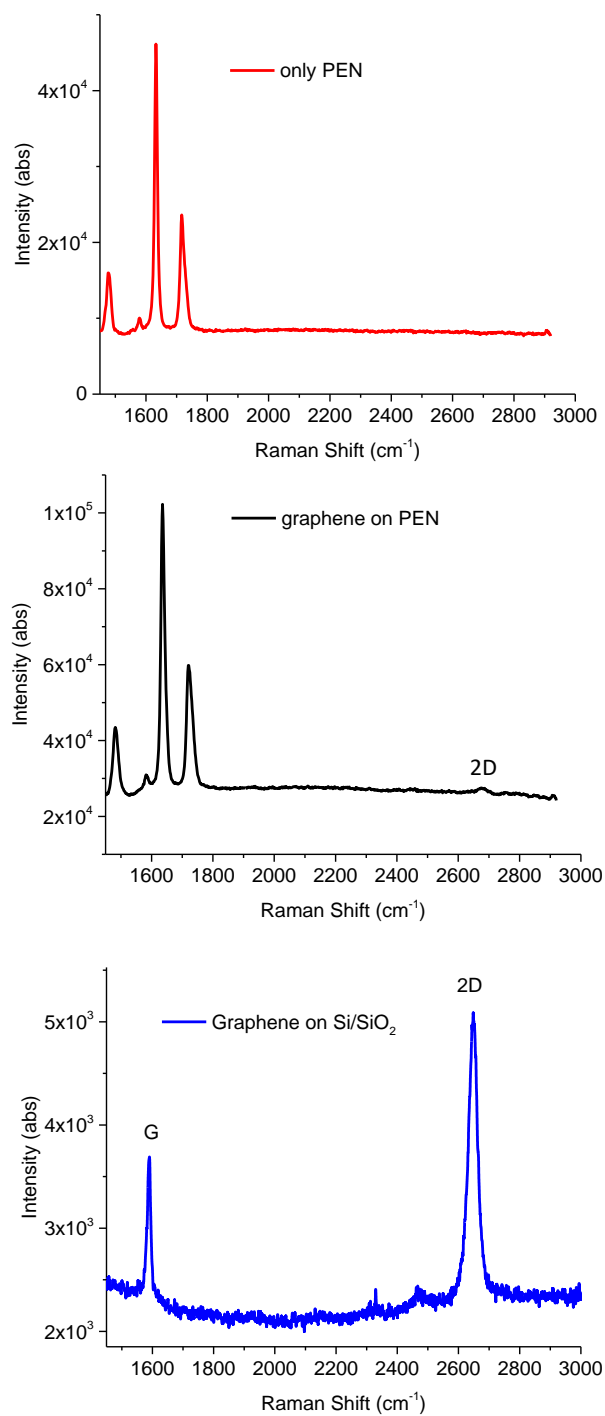

Fig. S1.2. Raman spectrum of PEN substrate, CVD graphene on PEN and Si/SiO$_2$ substrates.

In general, CVD graphene grown on the copper substrate is monolayer graphene due to the self-limiting growth process. It is not possible to characterize graphene over PEN using Raman spectroscopy due to the high background intensity (in Fig. S1.2). To assess the quality, we performed Raman spectroscopy of graphene over SiO$_2$, which showed an intensity ratio of 2D and G peak ~2 confirming a good quality of the single layer graphene.

## 2. Optimum resistance of the contacts

Fabricating electrodes with contact resistance ideal for spin injection is one of the prime challenges in realizing spin transport through graphene. Although it is possible to obtain low resistance contacts by directly depositing a ferromagnetic (FM) metal on graphene, such contacts are prone to contact-induced spin relaxation resulting in very weak on undetected spin signals. The migration of FM atoms into the graphene lattice can also lead to spin relaxation, which is why it is necessary to isolate the FM spin polarizer from the graphene layer. Conventionally, oxide barriers ($TiO_2$, MgO, $Al_2O_3$), and more recently, two-dimensional insulating crystals[1–3] have been employed to achieve efficient spin injection into graphene. Although tunneling-based electrical injection of spin-polarized electrons into graphene is efficient, such contacts possess significantly high resistance. While increasing the contact resistance can mitigate the problem of contact-induced spin relaxation, too high resistance values make the contacts unsuitable electrically. Considering these factors, the dimensions and channel resistance of the non-magnetic (NM) component of a spin valve, FM-NM-FM, the optimum range for efficient spin injection can be estimated using the Fert-Jeffrè model[4,5]. This model gives the value of calculated magnetoresistance (MR) as a function of the contact resistance of FM-NM junction. The Fig.4e in the main manuscript is based on this calculation non-local spin signal $\propto$ MR[4,5].

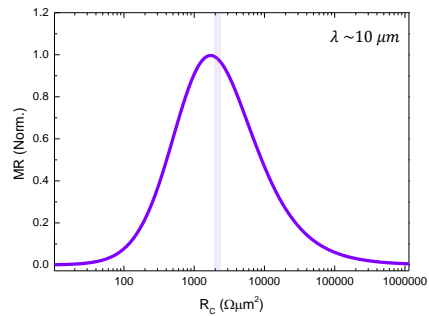

Fig. S2. **Calculated spin signal as a function of tunnel contact resistance**. Dependence of 2-terminal magnetoresistance (MR) on the contact interface resistance as calculated using the Fert-Jeffrè model[4,5] (calculated with device-1 electrical parameters).

The ideal window for optimum contact resistance for graphene in our case is shown in figure Fig. S2, where the variation of the two-terminal magnetoresistance for a graphene spin valve is plotted as a function of the contact resistance of the spin injection/detector electrodes. As shown in Fig. S2, by optimizing the titanium seed layer processing prior to depositing Co|Au, we were able to obtain the necessary contact resistance (measured by three-terminal measurements described in the main article) for spin signal observation. The obtained range of contact resistances in our device-1 is shown by the shaded region. Depending upon the variation in graphene sheet resistance, variation in device to device electrical parameters has been observed (as seen in device-2).

## 3. Carrier mobility measurement

There are several methods for estimating the carrier mobility and concentration in graphene. One of the most convenient ways for graphene devices on Si/SiO$_2$ substrates is to measure the field effect mobility from the conductivity of graphene channel recorded as a function of varying gate voltage. In Si/SiO$_2$ substrates, one can utilize the highly doped Si as the back gate electrode. The widely reported values of such field effect mobility are ~2000 cm$^2$V$^{-1}$s$^{-1}$ for both exfoliated and CVD graphene on Si/SiO$_2$ substrates. In the case of flexible PEN substrate, there is no such possibility to apply a back gate due to the insulating nature of the entire substrate. While it is possible to fabricate a top gate on graphene by depositing an oxide layer such as titanium dioxide or aluminum oxide, this would significantly dope graphene and alter its pristine electronic quality. To avoid such problems, we chose to perform Hall Effect experiments on several samples with varying square resistance from the same quality CVD graphene. In Fig. S3, we display the Hall voltage signals on three samples, measured in a standard Hall configuration (shown in the inset of the figure). From these measurements, we obtain a carrier concentration $n = \frac{I \partial B}{e |\partial V_H|}$ of ~10$^{12}$cm$^{-2}$ (0.9-2.04 × 10$^{12}$cm$^{-2}$) in several samples showing $R_{Sq}$ within the range of the resistance of spin transport devices. Assuming a mean carrier concentration ~1.5 × 10$^{12}$ cm$^{-2}$, we arrive at a carrier mobility $\mu = \frac{1}{R_{Sq}\, n\, e}$ ~ 10 000 cm$^2$V$^{-1}$s$^{-1}$ in our samples with low square resistance. Using the minimum carrier density ~10$^{12}$ cm$^{-2}$, yields a carrier mobility ~ 15 000 cm$^2$V$^{-1}$s$^{-1}$. Although it is not possible to obtain exact carrier concentration in the spin transport samples, the current approximation is a fair assumption[6] used for graphene samples on insulating substrates.

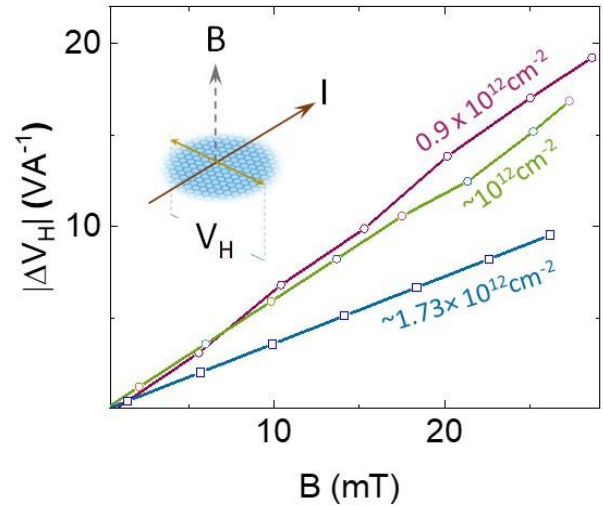

Fig. S3. **Hall measurements.** Hall voltage per unit current recorded as a function of magnetic field for three different samples. The inset shows the configuration of the Hall voltage measurement.

## 4. Atomic force microscopy (AFM) analysis of the topography of PEN substrate

Conventional Si/SiO$_2$ substrates have a mean roughness of ~0.3 nm, which leads to a significant reduction in the graphene electronic quality due to substrate-induced electron-hole puddles and scattering. The electronic quality of graphene can be enhanced several folds by suspending it or placing the graphene flakes on atomically flat hexagonal boron nitride micrometer size flakes. The best quality industrial grade PEN substrates used for flexible electronic applications have surface roughness ~1.3 nm, a roughness that is much greater than that of Si/SiO$_2$ substrates. In Fig. S4a and S4b, an AFM scan of the topography of PEN substrate is shown (the thickness axis is not in proportional to the x and y-axes). A typical line scan of the surface (Fig. S4c) exhibits a roughness width ranging from 50-100 nm with peak heights ~1.2 nm above the mean value. A 1-2 nm peak distributed over 50-100 nm (roughness width) has a nearly flat nature with a negligible slope. In addition to that, an inspection of the intra-peak roughness, (shown in Fig. S4d), reveals the roughness over the humps to be in the order of a few Å, which indicates that the PEN substrate is significantly smoother than what one would guess from the overall roughness value. Thus, the PEN surface is significantly smoother, could potentially lead to lower substrate-induced scattering of charge carriers.

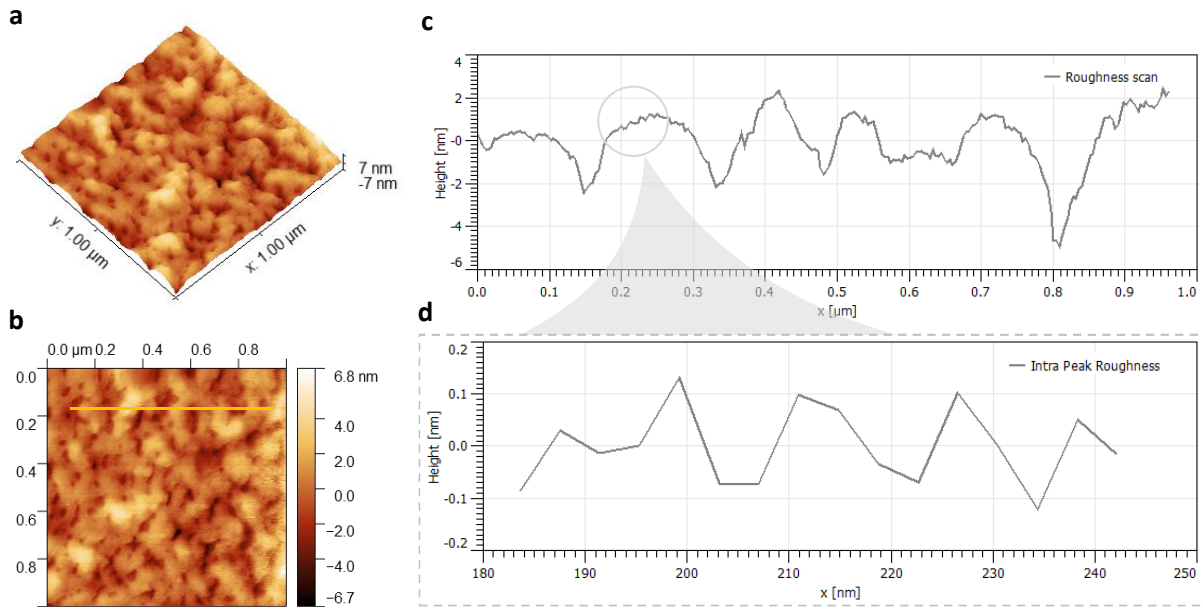

Fig. S4. **Atomic force microscopy of PEN substrate**. **a,** a three-dimensional view of a 1 μm x 1 μm scan of the PEN surface. **b,** Atomic force microscope image with line scan. **c,** Line scan taken over the AFM image of the PEN substrate. **d**, A zoomed scan profile of the intra-peak roughness.